\documentclass[prl,twocolumn,preprintnumbers,showpacs]{revtex4}
\usepackage{amsmath,amssymb,bm,epsfig,multirow}
\renewcommand\d{\partial}
\newcommand\x{{\bm{x}}}
\newcommand\y{{\bm{y}}}
\newcommand{\sect}[1]{\emph{#1.}---}

\begin{document}
\preprint{INT-PUB 07-51}

\title{Universal Fermi Gas with Two- and Three-Body Resonances}
\author{Yusuke~Nishida}
\author{Dam~Thanh~Son}
\author{Shina~Tan}
\affiliation{Institute for Nuclear Theory, University of Washington,
             Seattle, Washington 98195-1550, USA}

\begin{abstract}
We consider a Fermi gas with two components of different masses, with
the $s$-wave two-body interaction tuned to unitarity.  In the range of
mass ratio $8.62<M/m<13.6$, it is possible for a short-range interaction
between heavy fermions to produce a resonance in a three-body channel.
The resulting system is scale invariant and has universal properties,
and is very strongly interacting.  When $M/m$ is slightly above the
lower limit 8.62, the ground state energy of a 2:1 mixture of heavy and
light fermions is less than 2\% of the energy of a noninteracting gas
with the same number densities.  We derive exact relationships between
the pressures of the unitary Fermi gases with and without three-body
resonance when the mass ratio is close to the critical values of 8.62
and 13.6. Possible experimental realization with cold atoms in optical lattices
is discussed.
\end{abstract}

\date{February 27, 2008}

\maketitle

\sect{Introduction} Dilute Fermi gas at infinite scattering
length~\cite{Eagles,Leggett,Nozieres} has attracted considerable
attention recently.  The system can be realized in atomic traps using
the Fesh\-bach
resonance~\cite{OHara,Jin,Grimm,Ketterle,Thomas,Salomon}.
This system is called Fermi gas at unitarity and has interesting 
universal properties.  For example, the energy of the ground state of a
symmetric Fermi gas at unitarity is
  $E = \xi E_{\rm free}$,
where $E_{\rm free}$ is the energy of a noninteracting system with the
same density and $\xi$ is a dimensionless number.  Most recent
evaluations for $\xi$ yield $\xi\approx0.4$.

Other systems with universal properties that have been considered so
far are variations of this unitary Fermi gas.  One can consider
different densities of the two fermion
components~\cite{Ketterle-pol,Hulet-pol}, or different masses of the
two fermions, or both~\cite{SadeMelo-uneqm,Yip-uneqm}.  Here we show
that there exists a novel type of system with both two- and three-body
interactions fine-tuned to resonance.

\sect{Three-body wave function} In order to realize a three-body
resonance, it is necessary to allow the two components of the Fermi
gas to have different masses.  We shall denote the mass of the heavy
fermion as $M$ and of the light fermion as $m$.  The ratio $u=M/m$ is a
free parameter.

Let us recall what is known for three-body physics, using the
Born-Oppenheimer approximation at first~\cite{Fonseca:1978kj}, but then
proceed to the exact result.  Consider a three-body system of two heavy
and one light fermions.  If $M\gg m$ one can use the adiabatic
approximation, separating the wave function into the fast part of the
light fermion and the slow part of the heavy fermions,
\begin{equation}
  \psi(\x_1,\x_2,\y) = \psi_{\rm slow}(\x_1,\x_2) 
  \psi_{\rm fast}(\y;\x_1,\x_2).
\end{equation}
The function $\psi_{\rm fast}(\y;\x_1,\x_2)$ satisfies the
Schr\"odinger equation of one particle moving in the field of two
static centers:
\begin{equation}
  -\frac{\hbar^2}{2m}\nabla^2_\y\psi_{\rm fast}(\y;\x_1,\x_2)=
   E(\x_1,\x_2)\psi_{\rm fast}(\y).
\end{equation}
The resonant two-body interaction implies that the asymptotics of
$\psi_{\rm fast}(\y)$ when $\y\to\x_i$, $i=1,2$ is
$|\y-\x_i|^{-1}+O(|\y-\x_i|)$.  One finds one negative energy level
with 
\begin{equation}
  E(R)= -\frac{C^2\hbar^2}{2mR^2}, \quad R=|\x_1-\x_2|,
\end{equation}
where $C=0.5671\ldots$ 
is the solution to equation $e^{-x}=x$.  We
then regard $E(R)$ as the potential energy between the two heavy
fermions.  Since it is a $R^{-2}$ potential, the wave function
$\psi_{\rm slow}$ near the origin behaves as
$R^{\alpha}Y_{lm}(\theta,\phi)$, where $\alpha$ is a solution to the
equation
\begin{equation}\label{alpha}
  \alpha(\alpha+1) = l(l+1) - \frac{C^2 u}2.
\end{equation}
Because of the Fermi statistics $l$ is odd.   When the right-hand side
of Eq.~(\ref{alpha}) is greater than $-\tfrac14$, there are two real
solutions $\alpha_+$ and $\alpha_-$, relating to each other by
$\alpha_++\alpha_-=-1$ ($\alpha_+>-1/2>\alpha_-$).  If the right-hand
side is less than $-\tfrac14$, which is the case for $l=1$ and $u>u_{\rm
max}\approx14.0$, then $\alpha$ is complex.  This corresponds to the
Efimov effect, where the system develops deep bound states in the
zero-range limit, which break scale invariance.

In the former case, $u<u_{\rm max}$, there are two further
possibilities.  If $\alpha_+>\tfrac12$ and $\alpha_-<-\tfrac32$, then the
solution $\psi_{\rm slow}\sim R^{\alpha_-}$ is not normalizable at the
origin.  Therefore, the only consistent boundary condition for
$\psi_{\rm slow}$ at the origin is $\psi_{\rm slow}\sim R^{\alpha_+}$.
However, when $\alpha_+<\tfrac12$, $\alpha_->-\tfrac32$, both asymptotics
$R^{\alpha_\pm}$ are normalizable at the origin.  Therefore,
generically $\psi_{\rm slow}$ will behave as
\begin{equation}
  \psi_{\rm slow}(R,\theta,\phi) = (c_+ R^{\alpha_+} + c_-
  R^{\alpha_-}) Y_{lm}(\theta,\phi),
\end{equation}
where the ratio $c_+/c_-$ is determined by the interaction.  For a
generic interaction $c_+/c_-$, which has dimension
$[\textrm{length}]^{\alpha_--\alpha_+}$, will be of order
$r_0^{\alpha_--\alpha_+}$, where $r_0$ is the characteristic
interaction range.  That means that at distances $R\gg r_0$ the
wave function behaves like $R^{\alpha_+}$.  However, by including an
attractive force between the heavy fermions (in addition to the
effective attraction induced by the light fermion), and fine-tuning its
strength, one can achieve the situation where only the coefficient of
the $R^{\alpha_-}$ piece is very large.  This corresponds to a
three-body resonance, and $c_+=0$ corresponds to this resonance being
at threshold. For $l=1$, this is possible when $u>u_{\rm
min}\approx7.77$. To verify our point, we considered a model potential
$
V(R)=({\hbar^2}/{M})\left({\beta_1}/{R^{12}}-{\beta_2}/{R^6}\right)
$
between the heavy fermions. Within the Born-Oppenheimer approximation, 
when $\beta_2/\beta_1^{2/5}\approx6.5+2.9\alpha_+$
($-\tfrac12\le\alpha_+\le\tfrac12$), which is not a strong enough
attraction to produce a two-body resonance, the desired three-body
resonance is achieved.

The Born-Oppenheimer approximation employed above is not entirely
justified, because the obtained values of $u_{\rm min}$ and $u_{\rm
max}$ are not parametrically large.  The conclusions, however, are
confirmed by an exact treatment of the three-body problem.  The wave
function $\Psi(\x_1,\x_2,\y)$ satisfies the free Schr\"odinger
equation with the two-body boundary condition when $|\x_i-\y|\to0$.  In
hyperspherical coordinates, its behavior is
\begin{equation}
  \Psi = R^\gamma f_l(\Omega),
\end{equation}
where $R$ is the hyperradius and $\Omega$ denotes the hyperangular
variables, and $\gamma$ is related to $\alpha$ by
$\gamma=\alpha-\tfrac32$.  For the $l=1$ channel, the power $\gamma$ is
determined from the equation~\cite{Petrov}
\begin{multline}\label{gamma}
  \frac{ \cos\bigl[(\gamma{+}1)\arccos\frac{u}{u+1}\bigr]
        -\cos\bigl[(\gamma{+}1)\arccos\frac{-u}{u+1}\bigr]}
  {(\gamma+1)\sin\pi\gamma}\\
  +\frac{ \cos\bigl[(\gamma{+}3)\arccos\frac{u}{u+1}\bigr]
        -\cos\bigl[(\gamma{+}3)\arccos\frac{-u}{u+1}\bigr]}
  {(\gamma+3)\sin\pi\gamma} \\
  = - \frac{2u^2}{(u+1)^3}\sqrt{2u+1}.
\end{multline}
Solving this equation, we find that $u_{\rm max}\approx13.6$ 
and
$u_{\rm min}\approx8.62$, 
which somewhat differ from the values obtained in the adiabatic
approximation.  At the qualitative level, however, the physics is the
same: for the mass ratio between these two values, one can fine-tune one
three-body channel to resonance.  Note that the mass ratio $u_{\rm
min}\approx8.62$ played a special role in the discussion of three-body
recombination in Ref.~\onlinecite{Petrov}.

\sect{Many-body physics} We now discuss the ground state energy of the
many-body system with both two- and three-body resonances.  The first
question is whether the system is stable toward collapse.  Such a
collapse can happen if there exists an Efimov state of four or more
particles for the mass ratio between 8.62 and 13.6.  This would be
very unusual, as it is before the three-body Efimov state appears.
Consider four-body states.  In the Born-Oppenheimer approximation,
there is no Efimov state of two heavy and two light fermions, since
there is only one bound state in the Schr\"odinger equation with two
static centers.  The case of three heavy and one light fermion is more
subtle, but here the three centrifugal potentials between the heavy
fermions make the Efimov effect less likely.  We shall proceed under
the assumption that there is no Efimov effect for any number of heavy
and light particles, and the system is stable.

Let us define $\xi$ as the ratio of the ground state energy $E$ to the
ground state energy $E_{\rm free}$ of the noninteracting system at the
same densities of the heavy and the light components, $n_h$ and $n_l$,
respectively,
\begin{equation}
  E(n_h, n_l) = \xi E_{\rm free}(n_h, n_l).
\end{equation}
Since the system has no intrinsic dimensionful parameter, $\xi$ can
depend only on two variables: the mass ratio $M/m$, and the relative
abundance of the two components.  For the latter, we define
\begin{equation}
  x = \frac{n_l}{n_h+n_l}.
\end{equation}

In general, the problem of finding $\xi(M/m,x)$ is difficult, since
there is no small parameter in the problem.  We shall concentrate on
the case when $M/m$ is slightly larger than the lower critical value
8.62.  In this case, the normalization integral for the three-body wave
function is dominated at small distances.  Thus, one can think about the
three-body system as a localized bound state.  The situation is
analogous to the behavior of the two-body state at resonance near four
spatial dimensions.  As the number of spatial dimensions $d$ approaches
four from below, the two-body wave function becomes more and more
dominated by short distances, which has led Nussinov and Nussinov to
suggest that $\xi\to0$
in this limit~\cite{Nussinov}.  This was confirmed by the $\epsilon$
expansion near $d=4$~\cite{Nishida:2006br}.

The view that the three-body system behaves as a tightly bound state
is confirmed by its behavior in an isotropic harmonic potential.  It
is known that each exponent $\gamma$, determined from solving the
free-space Schr\"odinger equation, is related to an energy level of 
the three particles in an isotropic harmonic potential~\cite{Shina},
\begin{equation}\label{level}
  E_0 = \left(\frac 92 + \gamma\right)\hbar\omega,
\end{equation}
where $\omega$ is the oscillator frequency.  When $\gamma\to-3$, 
the energy approaches $\tfrac32\hbar\omega$.  This is consistent with
the picture of the trimer being a localized bound state of all three
particles.  We emphasize that the three-body state is not a real bound
state with a finite binding energy.  In particular, in the harmonic
trap there are breathing modes, corresponding to a ladder of energy 
excitations $E_0+2n\hbar\omega$, $n=1,2,\ldots$ \cite{WernerCastin},
that are absent for the single particle.  However, when the trimer is in
the ground state with respect to the breathing mode, it behaves as a
pointlike particle.

Having in mind this picture of trimers as objects localized in space,
let us put an upper bound on the value of $\xi$ at $x=\tfrac13$, and
$u=8.62+\epsilon$.  In this case one can combine all heavy and light
particles (whose relative number is 2:1) into trimers, and construct a
dilute gas of such trimers.  This gas will be a weakly interacting Fermi
gas of trimers with three possible polarizations (the trimer has $l=1$).
Its energy density is therefore
\begin{equation}
  \frac{E_{\rm trimer}}V = c \frac {n_l^{5/3}}{3^{2/3}(2M+m)},
\end{equation}
where $c=\frac{3}{10}(6\pi^2)^{2/3}\hbar^2$ and $n_l$ is the density of
the light fermions.  On the other hand, if one turns off the
interaction, the system becomes a Fermi gas of light fermions with
density $n_l$ plus a Fermi gas of heavy fermions with density
$n_h=2n_l$.  The energy density of this gas is
\begin{equation}
  \frac{E_{\rm free}}V = c \frac{n_l^{5/3}}{m} + c\frac{(2n_l)^{5/3}}M.
\end{equation}
Taking the ratio we find
\begin{equation}
  \frac{E_{\rm trimer}}{E_{\rm free}} = 
  \frac{u}{3^{2/3}(u+2^{5/3})(2u+1)} \approx 1.93\times 10^{-2}.
\end{equation}
It is clear that the trimer gas energy is not the lowest ground
state energy and hence is only an upper bound on the latter.  For
example, a trimer with energy near the Fermi energy can decay into
its constituents (two heavy and one light fermions) since the binding
energy is zero, and this decay can happen until chemical equilibrium
is reached.  The true ground state is therefore a mixture of the trimer
gas with a gas of constituent fermions.  Because of the strong
interactions between constituent fermions, we do not have an exact
formula for the ground state energy.
The upper bound, on the other hand, implies that the system is very
strongly interacting: the ground state energy is less than 2\% of the
energy of the noninteracting system:
$\xi\bigl(u\to u_{\rm min},\tfrac13\bigr) < 1.93\times 10^{-2}$.
This can be contrasted with the case of symmetric unitary Fermi gas with
equal masses, where $\xi\approx0.4$.

\begin{figure}[t]
 \includegraphics[width=0.4\textwidth,clip]{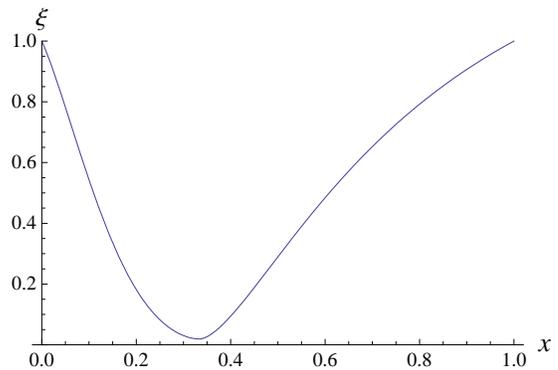}
 \caption{The variational bound on $\xi$ as a function of the abundance
of light fermions $x$ at $M/m=8.62+\epsilon$.
 \label{fig:bound}}
\vspace{-1ex}
\end{figure}

For a general value of $x$, we can put an upper bound on $\xi$ by
considering the state where the maximal number of trimers is bound and
form a trimer Fermi gas, while the remainder (heavy fermions for
$x<\tfrac13$ and light fermions for $x>\tfrac13$) form another free
Fermi gas.  The resulting bound is
plotted
in Fig.~\ref{fig:bound}.
One sees that the variational bound is lowest when $x$ is close to
$1/3$. 
From the point of view of the ground state energy, the 2:1 mixture of
heavy and light fermions is therefore the most interesting.

\sect{Exact relationships near $u=$ 8.62 and 13.6} 
For mass ratios $u$ between 8.62 and 13.6, therefore, there exist two
scale invariant regimes, which give rise to unitary Fermi gases with and
without three-body resonance.  We now show that near the lower critical
value $u_{\rm min}$ there exists an exact relationship between the
pressures of a unitary Fermi gas with three-body resonance, $P$, and a
unitary Fermi gas without three-body resonance, $P_0$, at the same
chemical potentials of light and heavy fermions, $\mu_h$ and $\mu_l$.  
Namely,
\begin{multline}\label{P}
  \lim_{u\to u_{\rm min}} [P(\mu_l,\mu_h) - P_0(\mu_l,\mu_h)]\\
  =  \frac{[2(2M+m)]^{3/2}}{5\pi^2\hbar^3} (2\mu_h + \mu_l)^{5/2}.
\end{multline}
We emphasize here that we have in mind the limit when for each mass
ratio $u$ we take the zero-range limit, compute the pressures $P$
and $P_0$, and then take the limit $u\to u_{\rm min}$.  In general
the limit $u\to u_{\rm min}$ and the zero-range limit do not commute.

The right-hand side of Eq.~(\ref{P}) is the pressure of an ideal Fermi
gas of trimers, with $l=1$, at chemical potential $2\mu_h+\mu_l$. 
Physically, Eq.~(\ref{P}) means that at $u\to u_{\rm min}$ the unitary
Fermi gas with three-body resonance consists of two parts: a trimer gas
and a Fermi gas of the fermionic constituents outside the trimers.
These two parts interact weakly with each other, but are in chemical
equilibrium.

When $x=\tfrac13$, there is another simple relationship between the
ground state energies with and without three-body resonance.  $\xi$ of
the Fermi gas with three-body resonance is determined by the analogously
defined quantity of the unitary Fermi gas without three-body resonance
$\xi_0$ via
\begin{equation}\label{xi}
 \xi^{-3/2} = \xi_0^{-3/2} + 
   374.
 \quad ( x=\tfrac13, ~ u\to u_{\rm min}). 
\end{equation}

We have argued above that the trimers decouple at $u_{\min}$.  What is
less trivial is that the fermions outside the trimers contribute the
same pressure as a Fermi gas without three-body resonance.  This seems
strange: the three-body wave functions satisfy different boundary
conditions at small distances ($R^{\gamma_-}$ vs. $R^{\gamma_+}$) in
the two cases.  To see that it is the case, consider the three-body
wave function of three particles at some positive energy.  The radial
part of the Schr\"odinger equation reads
\begin{equation}
  \frac{\d^2\psi}{\d R^2} + \frac 5R \frac{\d\psi}{\d R} 
  - \frac{\gamma(\gamma+4)}{R^2}\psi = -k^2\psi(R).
\end{equation}
This equation has two solutions:
\begin{equation}\label{Jpmnu}
  \psi(R) \sim R^{-2} J_{\pm\nu}(kR), \quad \nu=|\gamma+2|,
\end{equation}
where $J_{-\nu}$ corresponds to the case with three-body resonance,
and $J_\nu$ to that without the resonance.  The limit $u\to u_{\rm
min}$ corresponds to $\nu\to1$.  By using the relationship between
the Bessel functions,
\begin{equation}
  J_\nu(z)\cos(\nu\pi) - J_{-\nu}(z) = \sin(\nu\pi) Y_\nu(z),
\end{equation}
one sees that in the limit $\nu\to1$ the two wave functions in
Eq.~(\ref{Jpmnu}) are the same up to a sign, except for a very small
region near the origin, $kR\sim\sqrt{1-\nu}$.  This means that, for the
fermions outside the trimers, which have finite kinetic energy, the
resonant and nonresonant boundary conditions are equivalent.  This
leads to the relationships~(\ref{P}) and (\ref{xi}).

On the other hand, the limit $u\to u_{\rm max}$ corresponds to
$\nu\to0$, where two solutions in Eq.~(\ref{Jpmnu}) coincide.  Therefore
there is no distinction between Fermi gases with and without three-body
resonance and we obtain another exact relationship
\begin{equation}
  \lim_{u\to u_{\rm max}} [P(\mu_l,\mu_h) - P_0(\mu_l,\mu_h)] = 0.
\end{equation}

\sect{Other channels} So far we have considered only the $l=1$
three-body channel.  One may ask whether similar fine-tuning can be done
in other channels.  To answer this question, one has to compute the
exponents $\gamma_+$ in other channels and look for the mass ratio $u$
where it crosses $-1$.  This crossing happens only for odd $l$.  For
$l=3$ it occurs at $u=70.1$, 
and the value of the mass ratio increases with increasing $l$.
For these large mass ratios, the Efimov effect has already taken hold
in the $l=1$ channel, and the system cannot be scale-invariant.  We
conclude that scale invariance with resonant three-body interaction
can be achieved only in the $l=1$ channel.

\sect{Experimental realization in a lattice}
Fermions of species $c$ and $b$ on a cubic lattice with the Hamiltonian 
\begin{multline}\label{lattice}
  H = -\sum_{<ij>}(t_h c_i^\dagger c_j+t_l b_i^\dagger b_j)
      + U \sum_i c_i^\dagger b_i^\dagger b_i c_i\\
      +\frac{1}{2} W \sum_{<ij>} c_i^\dagger c_j^\dagger c_j c_i
\end{multline}
(where $\sum_{\langle ij\rangle}$ extends over nearest-neighbor lattice
sites) approach the above universal Fermi gas limit when the average
number of fermions per lattice site $\ll1$, if parameters are tuned to
get $cb$ and $ccb$ resonances: $U/(t_h+t_l)
\approx-3.957$,
$8.62\le t_l/t_h\le13.6$, and numerically determined $W$ is shown in 
Tab.~\ref{tab:W} (currently we only know a lower bound for $W$ at
$t_l/t_h=13.6$). One can easily show that $p$-wave $cc$ resonance is not
reached until $W/t_h\approx-9.53$.
\begin{table}
\begin{tabular}{c|c|c|c|c|c|c|c}
$t_l/t_h$ & 8.62 & 9 & 10 & 11 & 12 & 13 & 13.6\\
\hline
\multirow{2}{*}{$W/t_h$}
&$-3.80$  &$-3.625$  &$-3.15$  &$-2.575$  &$-2.0$  &$-1.2$  &$>-0.4$ \\
&$\pm0.03$&$\pm0.025$&$\pm0.05$&$\pm0.075$&$\pm0.1$&$\pm0.1$&
\end{tabular}
\caption{\label{tab:W}Approximate values of $W/t_h$ for three-body
resonance on a lattice.} 
\end{table}

The model~\eqref{lattice} may be experimentally realized with cold atoms 
on an optical lattice.  If the $cb$ scattering length in free space is
small and negative, we can tune $cb$ to resonance by tuning the lattice
depth \cite{Zoller-geometric}. $t_l/t_h$ is also tunable.
Nearest-neighbor attraction between $c$ atoms may be mediated by a third
kind of atom having a much smaller effective mass than $c$ and $b$,
interacting with $c$ atoms with an effective scattering length $a_{c3}$: 
$a_{c3}<0$ and $\lvert a_{c3}\rvert$ slightly exceeds the lattice spacing.
Tuning $a_{c3}$ with a magnetic field near a $c3$ Feshbach resonance, 
we can tune the magnitude of $W$ in Eq.~\eqref{lattice}.
The lattice also dramatically increases the lifetime of the system 
by suppressing the three-body recombination rate \cite{Petrov-heteronuclear}.

\sect{Conclusion} We have argued that there exists a class of universal
Fermi gases with resonances in both two- and three-body channels.  This
system remains strongly coupled in the whole range of mass ratio,
$8.62<M/m<13.6$, where it exists.  We have discussed the possibility of
realizing these gases with cold atoms in an optical lattice.  The
relevant lattice model, Eq.~\eqref{lattice}, may also be studied with
quantum Monte-Carlo simulations.

We thank Soon Yong Chang for discussions. 
Y.\,N. is supported, in part, by JSPS Postdoctoral Program for
Research Abroad.  This work is supported, in part, by DOE Grant No.\
DE-FG02-00ER41132.

\end{document}